\documentclass[11pt]{article}
\usepackage[utf8]{inputenc}
\usepackage[T1]{fontenc}
\usepackage{lmodern}
\usepackage{amsmath, amssymb, amsthm}
\usepackage{graphicx}
\usepackage{hyperref}
\usepackage{cleveref}
\usepackage{geometry}
\usepackage{authblk} 
\usepackage{float}
\usepackage{placeins}
\usepackage{bm}
\usepackage{bbm}
\usepackage{xcolor}

\title{Multi-phase-field Models of Biological Tissues}
\author[1]{Siavash Monfared\thanks{siavash.monfared@nbi.ku.dk}}
\author[1]{Aleksandra Arda\v{s}eva\thanks{aleksandra.ardaseva@nbi.ku.dk}}
\author[1]{Amin Doostmohammadi\thanks{doostmohammadi@nbi.ku.dk}}
\affil[1]{Niels Bohr Institute, University of Copenhagen, Copenhagen, Denmark, 2100 }
\date{}
\begin{document}

\maketitle

\begin{abstract}
Understanding how cells coordinate their behaviors to produce large-scale patterns and functions is central to deciphering biological processes ranging from tissue development and regeneration to cancer progression and morphogenesis. Despite advances in imaging and mechanical characterization, the role of physical forces in collective cell dynamics remains incompletely understood. Physics-based models are essential for complementing experimental data, offering access to high-resolution spatiotemporal fields and enabling mechanistic insights into complex multicellular systems. This review focuses on dense, soft tissues, where the mechanical deformation of one cell drives re-organization of its neighbors, giving rise to emergent behaviors such as orientational order and long-range force transmission. The multi-phase-field model provides a powerful and versatile framework to investigate such systems, bridging biological phenomena and the non-equilibrium physics of active matter. We discuss the theoretical foundations of the model and its applications to a range of biological contexts, including cell migration, heterogeneous populations, confined geometries, and metastasis. We also emphasize the integration of simulations with experimental data, highlighting how this approach is reshaping our understanding of tissue mechanics, collective order, and force transmission. Finally, we outline current trends and future challenges in applying multi-phase-field models to biology and soft matter physics.
\end{abstract}

\section{Introduction}
Concluding the case for his theory of evolution, Charles Darwin wrote~\cite{darwin1859origin} in \textit{On the Origin of Species}, `from so simple a beginning endless forms most beautiful and most wonderful have been, and are being evolved'. Darwin was discussing the evolution of diverse lifeforms from one ancestral organism, but just as magical is the development of a complex multicellular organism from a single parent cell. This process is a highly orchestrated affair that cells conduct collectively without any central guidance, creating intricate dynamic patterns essential to development and regeneration. However, despite the remarkable advances in imaging spatiotemporal dynamics of cell collectives and techniques for mechanical characterization both \textit{in vivo} and \textit{in vitro} ~\cite{Maskarinec2009, Nier2016, Mongera2018, Latorre2018, Mo2024}, the role of physical forces on biological functions remains poorly understood. In this vein, physics-based models play a critical role in complementing experiments, providing access to high-resolution spatiotemporal fields in three dimensions. A particularly important but challenging to study is a dense, soft multicellular system, such as tissues, where mechanical deformation of one cell necessitates re-organization of neighboring cells. It is within this context that a multi-phase-field model shines, offering a rich physics-based framework to advance our understanding of biological systems while providing a robust playground for non-equilibrium statistical physics. 

This review is organized as follows: We begin with an introduction to physical models for cell collectives, highlighting the importance of understanding collective cell behavior in development and regeneration. Next, we delve into the multi-phase-field model, discussing its foundational aspects, including time-scale separability, free energy functionals for passive interactions, and non-equilibrium behavior driven by active processes. We then explore the physics of active matter, focusing on collective self-organization, topological defects, active stress chains, and the emergence of biological phenomena. In the subsequent section, we bridge biological physics with experimental integration, covering quantitative modeling informed by experiments, cell migration in 2D and 3D, heterogeneous cell populations, and confined systems. Finally, we conclude with a discussion of current trends, the importance of multi-phase-field models in biological and physics research, and future challenges. 

Due to the ability to incorporate changes in shape, cell-cell, and cell-substrate interactions, as well as subcellular details, phase-field models have provided valuable theoretical predictions of the biophysics of single cells that have been tested experimentally (reviewed in \cite{aranson2016physical}). Throughout, we emphasize the role of physics-based models in advancing our understanding of multicellular systems, while noting that this review does not address single-cell modeling, molecular-level dynamics, or non-biological applications of multi-phase-field models. 

\subsection{Physical models for cell collectives}
Multicellular assemblies are integral to numerous biological processes, including tissue formation~\cite{Mongera2018,Petridou_2019}, wound healing~\cite{Tetley_2019,Jain_2020}, and cancer metastasis~\cite{Grosser_2021,Blauth_2021}. These assemblies, composed of interacting cells, exhibit complex behaviors essential for the proper functioning of biological systems. Understanding and modeling these behaviors are crucial to the advancement of fields such as developmental biology, regenerative medicine \cite{Vining2017,Chiou2018}, and cancer research \cite{Tse2011,Amos2021}.

One of the primary challenges in studying multicellular assemblies is the inherent complexity of the multi-body physics problem. Each cell within an assembly is constantly changing shape and dynamically interacting with its neighbors and the extracellular environments. These interactions are highly nonlinear and multiscale in nature, including adhesion, repulsion, and mechanical stress due to deformation. Additionally, cells can produce work, migrate, divide, and extrude, further complicating the modeling process.

Although this review focuses on the latest advances in multi-phase-field models and their application to living cells, it is worth briefly discussing other types of approaches. A more comprehensive overview can be found elsewhere \cite{aranson2016physical,camley2017physical,Moure2019,Alert2020}: (a) Active network models include vertex-based \cite{Farhadifar2007} and Voronoi-based \cite{Bi2016} approaches that, in most cases, represent a confluent layer through geometric description of each cell and the associated effective energy functional for the work required to deform that cell. 
Two-dimensional active network models have played a key role in explaining some of the fundamental biological observations \cite{Park2015,Cislo2023} with recent advances extending the model into the third dimension \cite{Okuda2015} and more complex force generation mechanisms \cite{Rozman2024}. (b) Particle-based models for living cells are based on granular physics concepts. These models have been instrumental in studying collective motion emerging from local interactions and force generation mechanisms \cite{Henkes2011,Basan2013,Smeets2016}. However, this approach cannot capture cell shape changes (deformation), which is critical for modeling dense, deformable multicellular assemblies. (c) Cellular Potts models \cite{Graner1992} represent another paradigm for modeling cellular systems where each cell is defined as a spin collection. 
These models can provide a detailed description of cell shapes. However, these shapes can be affected by artifacts due to the lattice choice. Furthermore, shape fluctuations depend on a temperature parameter that cannot be directly calibrated against experimental measurements. (d) Continuum models 
\cite{Toner1995,Toner1998,simha2002hydrodynamic,Marchetti2013,doostmohammadi2018active,Maitra2020,armengol2023epithelia} typically leverage hydrodynamics and have been instrumental in modeling the behavior of cell collectives. However, they provide a coarse-grained description of an inherently discrete system and are thus unable to resolve individual cell shapes as well as local force, and density fluctuations that are now understood to be fundamental to many biological functions. 
Such insights are less likely to be obtained from mean-field descriptions of relevant fields.

The multi-phase-field model overcomes some of these limitations It can naturally handle deformable interfaces and efficiently scale to a large system in three dimensions, a necessity for understanding inherently three-dimensional biological processes such as embryogenesis and tumorigenesis. Furthermore, its discrete nature, at the scale of an individual cell, can capture large fluctuations while providing a robust way to model proliferation, genetic mutations and/or expressions. This framework is also able to represent an arbitrary level of confluence~\cite{lober2015collisions,marth2016collective,ardavseva2022bridging} and rather flexible in incorporating various passive and active interactions in a concise and consistent manner. As such, the multi-phase-field model provides a robust framework to establish an integrated understanding of living matter as an interplay between mechanical cues, genetics, and biochemical signals.

\section{Multi-phase-field model}
Early phase-field models for modeling cells focused on single and multi-component vesicles \cite{Biben2003,Lowengrub2009} and single cell morpho-dynamics \cite{Kockelkoren2003,Shao2010,Ziebert2011,Moure2017}. This was followed by models for cell collectives \cite{lober2015collisions,Palmieri2015,mueller2019emergence,loewe2020solid,Wenzel2021} including a number of three-dimensional multi-phase-field models that offer unprecedented insights into the physics of cell collectives \cite{Nonomura2012,MoreiraSoares2020,monfared2023mechanical,Kuang2023}.

In the multi-phase-field modeling approach, each cell is represented as an active droplet interacting with a surface. Generally in such a system, the drag due to the surface interactions dominates, leading to overdamped (translational) dynamics. Without loss of generality, consider a cellular monolayer that consists of $N$ cells on a rigid substrate with its surface normal $\vec{e}_{n}\left(=\vec{e}_{z}\right)=\vec{e}_{x}\times\vec{e}_{y}$ and periodic boundaries in both $\vec{e}_{x}$ and $\vec{e}_{y}$, where $\left(\vec{e}_{x},\vec{e}_{y},\vec{e}_{z}\right)$ constitute the global orthonormal basis. Each cell $i$ is represented by a three-dimensional phase-field $\phi_{i}=\phi_{i}\left(\vec{x},t\right)$. Interface dynamics is modeled via this auxiliary phase-field order parameter that varies smoothly between inside and outside of each cell domain. Such a construct eliminates the need for explicitly defining and tracking the interface spatiotemporally while providing the resolution necessary to resolve intercellular interactions. In turn, this leads to a highly scalable framework suitable for studying emerging behavior in large systems of interacting cells, such as tissues and organs. Furthermore, the free energy functional at the core of a phase-field model provides a versatile and elegant mathematical and thermodynamically consistent framework to model a wide range of physiochemical interactions.

\subsection{Time-scale separability: translational and relaxational dynamics}
Formation of an embryo \cite{Chan2019,Nelson2022}, tissue repair \cite{Kuipers2014,Ladoux2017}, and metastasis \cite{khalil_collective_2017,Padmanaban2019,van2023cell} emerge from collective interactions of living cells. This involves coordinated regulation of cell shape deformation and motion in space and time, without any central guidance, leading to an inherently multiscale process, ranging from molecular to multicellular length scales and spanning timescales from milliseconds to days \cite{beedle2023reversibility}.

One way to approach modeling the collective self-organization in living cells is to consider two sub-problems: (i) translational dynamics associated with cell movement and (ii) relaxational dynamics associated with cell deformation. Implicit in this approach is the assumption that the dynamics associated with each sub-problem satisfies time scale separability. This is indeed a justified assumption based on experimental characterizations of cell shape relaxation \cite{Balasubramaniam2025}. This paves the way to decouple two partial differential equations, one describing the (relatively) fast dynamics of relaxation and the other concerned with (relatively) slow dynamics of translation. For the translational dynamics, an overdamped Langevin type dynamics can be considered: 
\begin{equation}\label{dyn_trans}
    \xi\vec{v}_{i}\left(\vec{x},t\right)=\vec{F}^{\text{active}}_{i}\left(\vec{x},t\right)+\vec{F}^{\text{passive}}_{i}\left(\vec{x},t\right),
\end{equation}
where $\xi$ parameterizes substrate friction, here considered a constant but can depend on space and time, $\vec{F}^{\text{passive}}_{i}$, $\vec{F}^{\text{active}}_{i}$ and $\vec{v}_{i}$ represent passive forces, active forces and velocity for cell $i$, itself represented by $\phi_{i}$. On the other hand, the relaxational dynamics of the interface is described by a time-dependent Ginzburg-Landau model, also known as \textit{model A} dynamics in Hohenberg-Halperin classification scheme \cite{Hohenberg1977}, with an extra advective term: 
\begin{equation}\label{dyn_intf}
 \partial_{t}\phi_{i}+\vec{\nabla}\cdot\left(\vec{v}_{i}\phi_{i}\right)=-\Gamma\frac{\delta\mathcal{F}}{\delta\phi_{i}},\qquad i=1,...,N ,  
\end{equation}
where $\phi_{i}(\vec{x},t)$ is a scalar field representing the interface associated with cell $i$. $\Gamma$ is the mobility coefficient, affecting the relaxation time scale and $\mathcal{F}$ represents the free energy functional. Furthermore, we assume a nearly incompressible system, i.e. $\vec{\nabla}\cdot\vec{v}_{i}\approx 0$, and thus the advective term $\vec{\nabla}\cdot\left(\vec{v}_{i}\phi_{i}\right)\approx\vec{v}_{i}\cdot\vec{\nabla}\phi_{i}$ updates the field $\phi_{i}\left(\vec{x},t\right)$ for each time step and each cell $i$. 

\subsection{Free energy functional: passive interactions}
Generally, passive forces derive from the energy minimization principle obeying detailed balance and time-reversal symmetry. To this end, one can define passive forces acting on each cell, $\vec{F}^{\text{passive}}_{i}$, as follows
\begin{eqnarray}\label{eq:fpass}
    \vec{F}^{\text{passive}}_{i} = \int d\vec{x}\left(\sum_{i}^{N}\frac{\delta\mathcal{F}}{\delta\phi_{i}}\right)\vec{\nabla}\phi_{i}.
\end{eqnarray}
Eq. \eqref{eq:fpass} describes the forces in a system defined by a free energy functional, $\mathcal{F}=\sum_{i}^{N}\mathcal{F}_{i}$, defined below:  
\begin{eqnarray}
\label{Energy}
\mathcal{F}_{i} &=& \frac{\gamma^{i}}{\lambda}\int d\vec{x}\{4\phi_{i}^{2}\left(1-\phi_{i}\right)^{2}+\lambda^{2}\left(\vec{\nabla}\phi_{i}\right)^{2}\}\notag \\ 
 &+& \mu\left(1-\frac{1}{V_{0}}\int d\vec{x}\phi_{i}^{2}\right)^{2}+\sum_{j\neq i}\frac{\kappa_{\text{cc}}}{\lambda^2}\int d\vec{x}\phi_{i}^{2}\phi_{j}^{2}\notag\\
 &+& \sum_{j\neq i}\omega_{\text{cc}}^{i}\int d\vec{x}\vec{\nabla}\phi_{i}\cdot\vec{\nabla}\phi_{j}+\frac{\kappa_{\text{cs}}}{\lambda^2}\int d\vec{x}\phi_{i}^{2}\phi_{w}^{2}\notag\\
 &+& \omega_{\text{cs}}^{i}\int d\vec{x}\vec{\nabla}\phi_{i}\cdot\vec{\nabla}\phi_{w}.
\end{eqnarray}

The first term in the free energy functional, defined by Eq.\eqref{Energy}, encapsulates the intrinsic energy of cell $i$. In this instance, this is achieved by a symmetric double-well potential that establishes a preference for two distinct phases, i.e. inside and outside of a cell as well as a gradient term. It penalizes spatial variation in $\phi_{i}$, where parameter $\lambda$ sets the length scale associated with the diffusive interface. The second term in Eq. \eqref{Energy} enforces a volume constraint, ensuring that volume of cell $i$ does not deviate significantly from imposed volume, $V_{0}=(4/3)\pi R_{0}^{3}$, where $R_{0}$ is the initialized cell radius. The energy cost due to any deviation grows quadratically with strength $\mu$. The third term in Eq. \eqref{Energy} specifies repulsion with strength $\kappa_{\text{cc}}$ between cells $i$ and $j$ with a quadratic form based on overlapping phase-fields (Fig. \ref{fig1}a), making the energy cost more sensitive to the degree of overlap. In general, any positive even exponent suffices to avoid attraction due to very small negative phase-field values that may arise numerically. The fourth term in Eq. \eqref{Energy} captures cell-cell adhesion interactions of strength $\omega_{\text{cc}}^{i}$. This term contributes only when the gradients of phase-fields $i$ and $j$ overlap spatiotemporaly, given the gradients are only non-zero at the interface. This differs from the repulsion term (third term in Eq. \eqref{Energy}) that is based on the overlap of phase-fields and not their gradients. The second to last and last terms have a similar structure to the third and the fourth terms for cell-cell repulsion and adhesion but with different parametrization for strength, i.e. cell-substrate repulsion $\kappa_{\text{cs}}$ and cell-substrate adhesion $\omega_{\text{cs}}$. In this three-dimensional approach, cell heights emerge from collective interactions (Fig. \ref{fig1}b), making it suitable to study inherently three-dimensional biological processes such as cell extrusion. Furthermore, $\phi_{w}$ is a static phase-field that represents an arbitrarily defined three-dimensional matrix, such as a dense fibrous network \cite{MoreiraSoares2020} (Fig. \ref{fig2}a). This same construct can also account for a deformable matrix where $\phi_{w}=\phi_{w}\left(\vec{x},t\right)$ evolves due to interactions with each cell and/or an externally imposed dynamics. This formulation can capture heterogeneity in interactions such as $\gamma^{i}$ and $\omega_{\text{cc}}^{i}$, given the constraints imposed for stability are satisfied (see e.g. \cite{Palmieri2015}). Furthermore, the described multi-phase-field approach is primarily concerned with physical interactions in active matter. However, such a framework can benefit from further expansion to include chemical activity \cite{Zwicker2016,Bauermann2022}, typically captured by a Flory–Huggins type free energy functional; coupling mechanics with (bio-)chemistry.
\subsection{Non-equilibrium behavior: active drive}
Cell motility refers to the ability of a cell to move as it senses and reacts to its environment \cite{Abercrombie1954,Abercrombie1958}. This is achieved mainly by local injection of energy supplied by adenosine triphosphate (ATP) fueling two processes that work together to generate self-propulsive forces: (a) actin polymerization that results in the formation of protrusions at the cell front and (b) myosin-driven contractility that pulls a cells' rear inward, enabling forward movement \cite{Carlier1997,Pollard2009,Yam2007}.
 
Such local injection of energy breaks time-reversal symmetry and detailed balance, giving rise to fascinating collective phenomena at length scales much larger than a single cell \cite{OByrne2022}, including intricate and dynamic patterns critical for e.g. embryogenesis. To this end, multicellular assemblies are a textbook example of active matter, defined as natural or synthetic systems composed of entities that dissipate energy to perform mechanical work on themselves and their environment \cite{Ramaswamy2010,Marchetti2013,Bechinger2016}. As such, living cells can be viewed as a particular instance of active matter with the ability to proliferate, differentiate, and mutate.

In this vein, activity is not limited to cell motility. It can also manifest at a length scale larger than an individual cell, e.g. due to proliferation \cite{Doostmohammadi2015,Hallatschek2023} and/or the poroelastic nature of cells and their surrounding environments; breaking conservation of mass, volume, and number densities. In this review, we primarily focus on multi-phase-field models with cell motility as the only source of activity. However, this modeling framework can indeed extend to account for activity due to expansion, such as cell division and/or poroelasticity.

\begin{figure*}[htb!]
\centering
\includegraphics[width=1\linewidth]{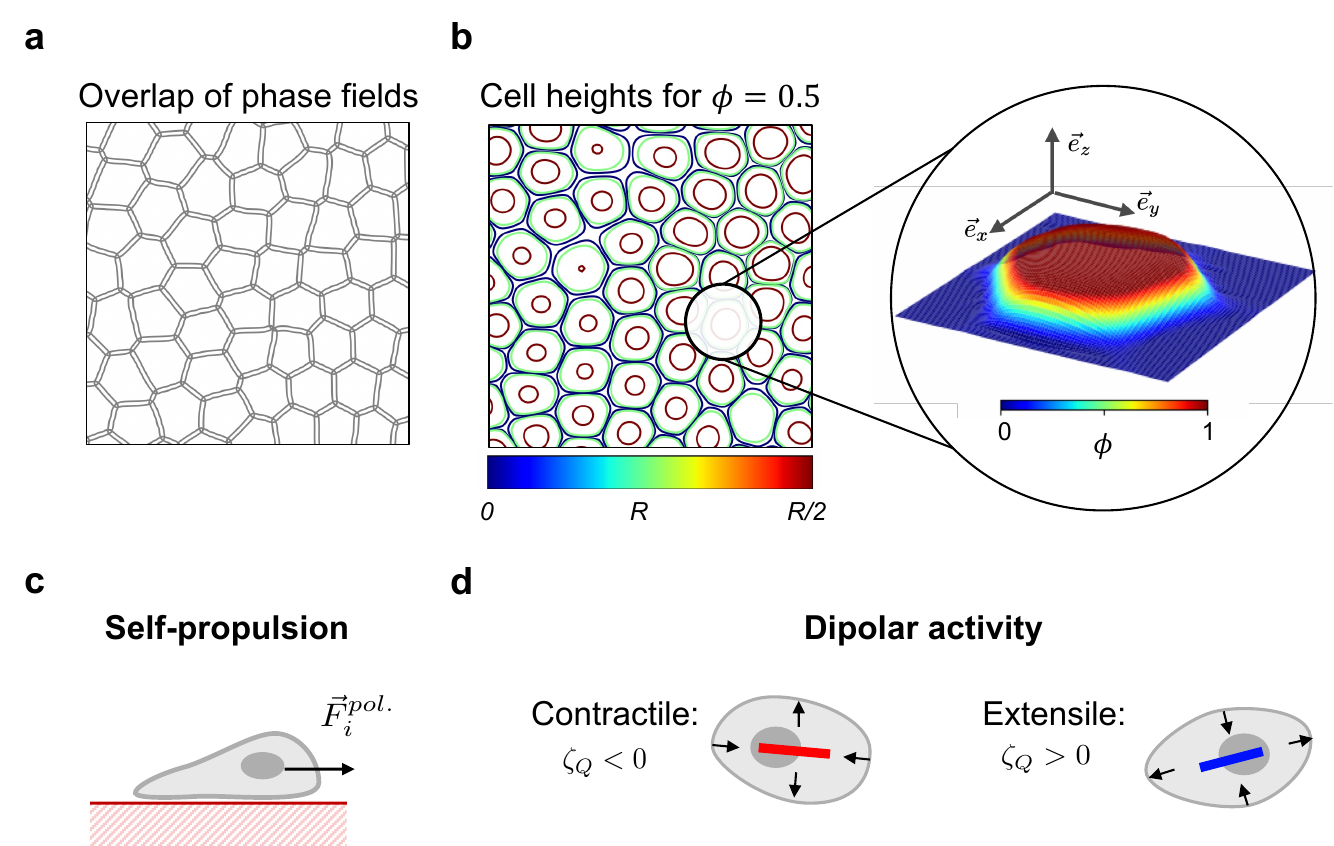}
\caption{\label{fig1} {\bf Multi-phase-field model.} a) Visualization of overlapping phase-fields for $\phi_i=0.1$. b) Visualization of cell heights and their fluctuations for a simulated 3D monolayer. The inset plot demonstrates a single scalar phase-field, $\phi$, in three dimensions. c) Schematic representation of self-propulsive force, $\vec{F}^{\text{pol}}_i$. d) Schematic representation of nematic active stresses for contractile and extensile systems. Red and blue lines represent the nematic director.}
\end{figure*}

In Eq.\eqref{dyn_trans}, the non-conservative active forces $\vec{F}^{\text{active}}_{i}$ are due to cell motility. A common physics-based model for cell motility uses the broken front-back symmetry from actin polarization. Generally, this manifests as a self-propulsion force that encodes a magnitude and a direction with some persistence to reflect directed motion (Fig. \ref{fig1}c). Two frequently employed models are: (a) Active Ornstein-Uhlenbeck Particles (AOUP) and (b) Active Brownian Particles (ABP) \cite{Solon2015,Martin2021,Fodor2022}. A major distinction between the two models is that both magnitude and direction are stochastic in AOUPs, while ABPs follow a stochastic direction with a constant speed. For example, consider the following ABP-type model: 
\begin{equation}\label{abp}
    \vec{F}_{i}^{\text{pol}}=\alpha_{i}\vec{p}_{i},
\end{equation}
where $\vec{p}_{i}=\left(\cos\theta_{i},\sin\theta_{i},0\right)$, a rank 1 tensor, is a front-back polarity vector attached to cell $i$, acting in-plane, and $\alpha_{i}$ parametrizes the strength of polarity, akin to a constant velocity in the ABP model. This further necessitates a description for polarization dynamics; e.g. defined here as a stochastic process that aligns the polarity of the cell to the direction of the total passive interaction force, $\vec{F}^{\text{passive}}_{i}$:
\begin{equation}\label{stp}
    \partial_{t}\theta_{i}=-\frac{1}{\tau_{\text{pol}}}\Delta\Theta_{i}+\sqrt{2D_{r}}\eta(t),
\end{equation}
where $\Delta\Theta_{i}$ is the angle between $\vec{p}_{i}$ and $\vec{F}^{\text{passive}}_{i}$. Furthermore, $\tau_{\text{pol}}$ sets the alignment time scale and $D_{r}$ represents rotational diffusivity, inversely related to persistent time. $\eta(t)$ denotes a standard Gaussian white noise with zero mean and unit variance, where $\langle\eta\left(t\right)\eta\left(t'\right)\rangle=\delta\left(t-t'\right)$. The model presented in Eqs. \eqref{abp} and \eqref{stp} agree well with the experimental characterization of collective motion of various cell cultures \cite{Peyret2019}.

An important component of cell motility hinges on myosin contractility. This contractile active force is typically captured by nematic stress that is proportional to the nematic order parameter, a rank 2 tensor \cite{simha2002hydrodynamic}:
\begin{eqnarray}\label{nematic_ten}
    \bm\sigma^{\text{nematic}}=\zeta_{\text{Q}}\sum_{i}\phi_{i}\bm Q_{i},
\end{eqnarray}
where $\zeta_{\text{Q}}$ represents the strength, contractile for $\zeta_{\text{Q}}<0$ and extensile for $\zeta_{\text{Q}} > 0$, and $\bm Q_{i}=2\left(\vec{n}_{i}\otimes\vec{n}_{i}-\frac{\bm{I}}{2}\vec{n}_{i}^{2}\right)$, where $\bm{I}$ is rank 2 identity tensor. This active nematic stress introduces a dipolar force density along cells' interfaces (Fig. \ref{fig1}d). The angle associated with $\vec{n}_{i}$ can follow its own dynamics \cite{ardavseva2022bridging} or it can be set equal to cell polarity, $\vec{n}_{i}=\vec{p}_{i}$, e.g. ensuring contractile stresses act in the same direction as protrusion formation. The nematic order along the direction $\vec{n}$ is head-tail symmetric, i.e. $\vec{n}\rightarrow-\vec{n}$. This implies that polar entities with broken front-back symmetry can self-organize into nematic ordering if on average they align together while not pointing to a particular direction.

Another rank 2 tensor construct for active drive focuses on the generation of intercellular force through the adherent junctions \cite{mueller2019emergence}. This approach utilizes a shape tensor, $\bm S_{i}$, to characterize shape anisotropy such that the direction associated with its largest eigenvalue captures the axis of elongation: 
\begin{eqnarray}\label{shape_ten}
    \bm\sigma^{\text{shape}}=\zeta_{\text{S}}\sum_{i}\phi_{i}\bm{S}_{i}
\end{eqnarray}
Here, $\zeta_{\text{S}}$ denotes the strength of activity, with the same sign convention as $\zeta_\text{Q}$, and $\bm S_{i}=-\int d\vec{x}\vec{\nabla}\phi^{T}\vec{\nabla}\phi$ is the traceless part of the negative of the structure tensor. Altogether, these active forces can be written as: 
\begin{eqnarray}\label{eq:factive}
    \vec{F}_{i}^{\text{active}}=\int d\vec{x}\left(\bm\sigma^{\text{nematic}}+\bm\sigma^{\text{shape}}\right)\cdot\vec{\nabla}\phi_{i}+\vec{F}_{i}^{\text{pol}}.
\end{eqnarray}
The interplay of emerging order and the nature of active force generation remains an intense area of research \cite{Maitra2020,Amiri2022,Cislo2023,armengol2023epithelia,Tan2022}, which we will discuss further in Section \ref{section:3}.

Lastly, using the definitions for active and passive forces (Eqs. \eqref{eq:factive} and \eqref{eq:fpass}, respectively) a rank 2 stress tensorial field can be constructed, providing access to in-plane and out-of-plane stress components. This is best achieved using a discrete stress definition consistent with those used in the granular physics and molecular dynamics communities \cite{Irving1950,Christoffersen1981}. Such a definition respects the discrete nature of a multi-phase-field model. For passive stresses another approach can be utilized based on Korteweg stresses for diffusive interfaces \cite{Kortweg1901}, commonly used in multiphase fluid mechanics for capillary stresses \cite{Anderson1997,Monfared2020}, most recently introduced in the context of a multi-phase-field model~\cite{chiang2024multiphase}.

\section{Multi-phase-field models and physics of active matter}\label{section:3}
\begin{figure*}[h!]
 \centering
\includegraphics[width=1\linewidth]{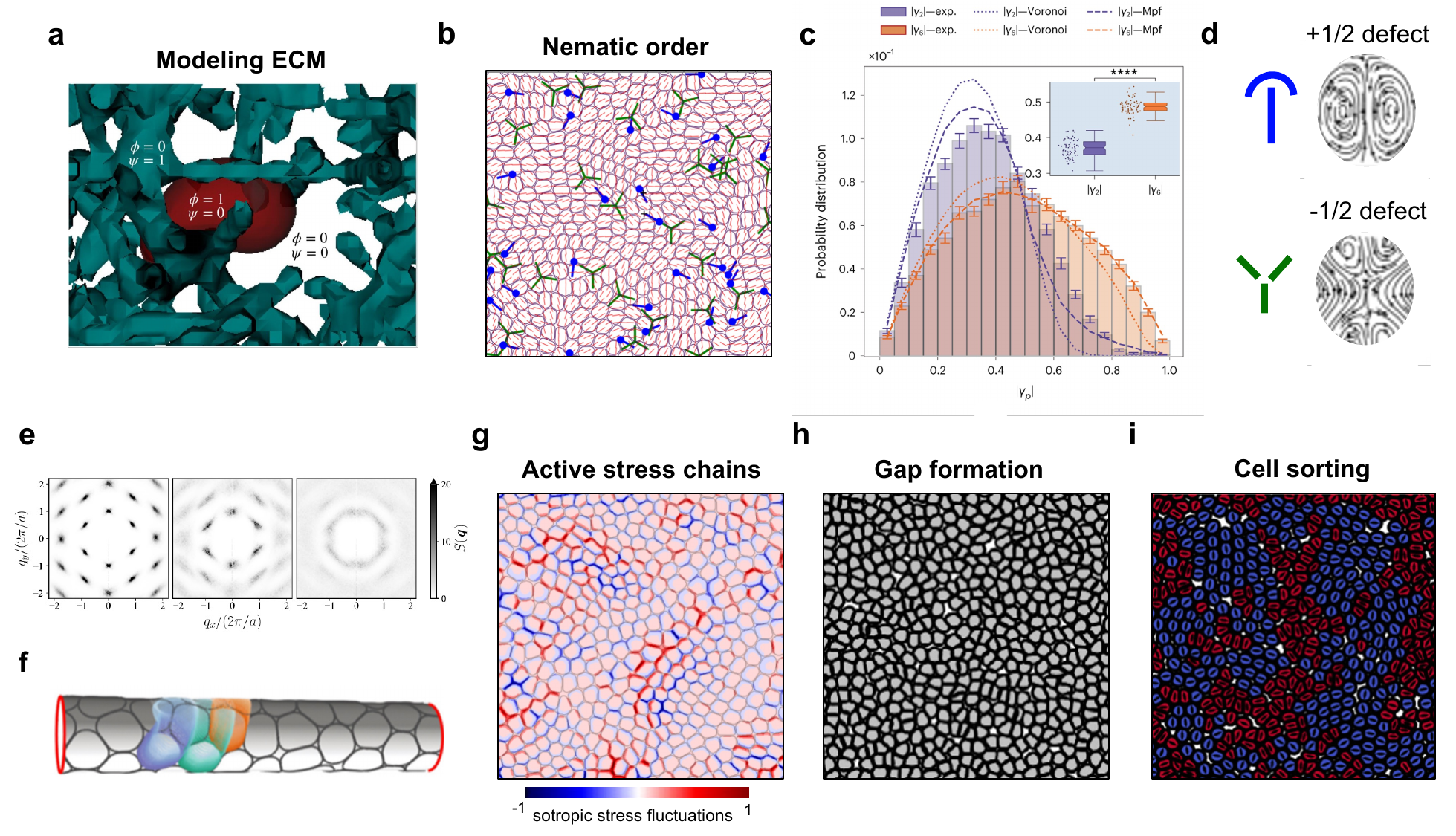}
\caption{\label{fig2} {\bf Multi-phase-field models and physics of active matter.}
a) Simulation setup where one phase-field ($\phi$) defines the cell, while the other ($\Psi$) -- extracellular matrix. Adapted from \cite{MoreiraSoares2020}. b) A coarse-grained director field plotted on top of cell interfaces ($\phi=0.5$) with +1/2 (filled circles with the line indicating orientation) and -1/2 (three connected lines with threefold symmetry) nematic defects. Adapted from \cite{monfared2023mechanical}. c) Probability distribution of experimental data (exp., bar plot) and Voronoi (dashed) and multi-phase-field (Mpf, dotted) simulations, with $p=2$ (purple) and $p=6$ (orange), demonstrating coexistence of nematic and hexatic order. Adapted from \cite{armengol2023epithelia}. d) Average flow-fields around +1/2 and -1/2 topological defects. Adapted from \cite{mueller2019emergence}. 
e) two-dimensional static structure factors for crystalline, hexatic, and isotropic liquid phases obtained from multi-phase-field model simulations. Adapted from \cite{Puggioni2025}. f) Snapshot of cell layer modeled on cylinder shape. Adapted from \cite{happel2024coordinated}. g) An example showing stress chains in an active monolayer. Adapted from \cite{monfared2024short}. h) Simulation snapshots of segregation in a 1:1 mixture of extensile (red) and contractile (blue) cells. Adapted from \cite{graham2024cell}. i) Snapshots of extensible cell monolayers demonstrating spontaneous formation of gaps. Adapted from \cite{ardavseva2022bridging}. }
\end{figure*}

From a physics perspective, the defining feature of a living matter is their non-equilibrium nature. In addition to providing significant biological insights, multi-phase-field models provide a rich playground to explore ideas in non-equilibrium statistical physics, in particular systems where constituent particles are capable of changing and adapting their shapes to the forces they experience. To this end, multi-phase-field models have been instrumental in exploring the collective motion and the dynamic organization of deformable and shape-changing active particles. A number of studies have leveraged multi-phase-field models to explore the emergence of order in active matter and its breakdown with significant consequences for living systems. This is discussed next, including exciting areas where multi-phase-field approach promises novel physics insights. 

\subsection{Collective self-organization and emergence of order}
Over the past two decades, extensive research has established a strong analogy between living matter -- at multiple length scales -- and active liquid crystals \cite{doostmohammadi2021physics,Bowick2022,Shankar2022}. This provides a rigorous theoretical framework to understand living matter and, in particular, dynamics of self-organization and collective behavior of cells and how it affects biological processes such as cancer metastasis \cite{Barberis2024}. Liquid crystals are an intermediate phase of matter between solids and liquids, characterized by a short-range translational order and a quasi-long-range orientational order \cite{gennes1993the,Nelson2004}. There are different phases of liquid crystals; generally named $p$-atic liquid crystals exhibiting $p$-fold rotational symmetry, i.e. symmetry with respect to rotations by $2\pi/p$. This forms the basis for the physics-based description of active force generation at the scale of a single cell and its interplay with the emergence of order at a much larger scale.

The order that emerges from self-organization in living cells governs the most fundamental biological processes such as development. Focusing on multicellular systems as opposed to subcellular ones, nematic order ($p=2$) and the associated half-integer topological defects have been reported in a number of cellular systems \cite{ardavseva2022topological}, including but not limited to Madine-Darby Canine Kidney (MDCK) cells \cite{sawtopological2017,balasubramaniam2021investigating}, human breast cancer cells \cite{PrezGonzlez2018} and neural progenitor stem cells \cite{kawaguchi2017topological}. Similarly, multi-phase-field models have shown the emergence of nematic order due to force generation mechanisms defined by rank 2 active stress tensors \cite{mueller2019emergence,ardavseva2022bridging} (Eqs. \eqref{shape_ten} and \eqref{nematic_ten}). Concurrently, the co-emergence of nematic (Fig. \ref{fig2}b) and hexatic orders from active polar forces, Eqs. \eqref{abp} and \eqref{stp}, has been captured by a multi-phase-field model \cite{monfared2023mechanical}. Furthermore, Armengol-Collado et al. have demonstrated co-existence of both nematic and hexatic orders in confluent cell layers depending on the length scales of interest: two-fold nematic ($p=2$) is dominant at larger scales, whereas six-fold hexatic ($p=6$) at smaller length scales \cite{armengol2023epithelia} (Fig. \ref{fig2}c). This is achieved by complementing {\it in vitro} experiments on confluent MDCK cells with multi-phase-field and self-propelled voronoi models. This hierarchical structure has also been recovered from a hydrodynamics description of confluent epithelial monolayers \cite{armengol2024hydrodynamics}, hinting at the generality of this approach. Recent experiments complemented with vertex-based modeling show how correlated cell division can lead to the emergence of tetratic ($p=4$) order \cite{Cislo2023}. Viewing multicellular assemblies as multiscale $p$-atic active liquid crystals provides a powerful framework for understanding the dynamics of self-organization and collective behavior of cells with important implications for $p-$atic defect dynamics that is crucial to some of the most fundamental biological processes.

\subsection{Collective self-organization and breakdown of order: topological defects}
Topological defects refer to singularities that disrupt field symmetry. In living cells, they can be formed due to geometric frustration, boundary conditions and/or activity responsible for driving the system away from thermodynamic equilibrium. One of the key features of active liquid crystalline systems is the continuous formation and annihilation of topological defects. The multi-phase-field models with shape-determined activity \cite{mueller2019emergence} and nematic dipolar activity \cite{ardavseva2022bridging} not only have reproduced the dynamics of topological defects but also accurately captured the flow fields and stresses around the defects (Fig.~\ref{fig2}d). This holds for both extensile and contractile systems. Furthermore, Zhang et al. incorporated polar forces that arise from cytoskeletal propulsion, observing a sharp transition from jammed to liquid states, as well as flocking \cite{zhang2020active}.

Topological defects have been associated with various biological functions, ranging from extrusion events in a cell monolayer \cite{sawtopological2017} to the development of limbs in animals \textit{Hydra} \cite{maroudas2021topological}. Addressing such problems and understanding the underlying physics require transitioning from two-dimensional modeling approaches to three-dimensional ones. For example, possibility of large overlaps being developed between the cells at the points of fivefold disclinations was predicted using a two-dimensional multi-phase-field model~\cite{loewe2020solid}, and a recent three-dimensional multi-phase-field model linked cell extrusions to both half-integer nematic defects and fivefold disclinations in the hexatic order with the strength of this link regulated by force transmission across the monolayer \cite{monfared2023mechanical}. Indeed, some confluent cell layers such as MDCK cells seem to behave as a multiscale $p-$atic liquid crystal \cite{armengol2023epithelia}. This concept requires a nuanced understanding of $p-$atic defect dynamics and its interplay with various biological processes. For example, a recent study links the unbinding of hexatic defects to the process of cell intercalation, providing a possible explanation for the extensile nature of in vitro epithelial layers \cite{Krommydas2024} through a process analogous to the KTHNY melting scenario \cite{Nelson1979}. A follow-up study uses a multi-phase-field model to explore this link further by focusing on the onset of collective cell migration and the two-dimensional defect-mediated melting; contrasting melting transition in active matter against its passive counterpart \cite{Puggioni2025} (Fig.~\ref{fig2}e). In this vein, the emergence of both nematic and hexatic orders in confluent/near confluent active monolayers are linked to cellular geometry informed by intercellular friction and motility \cite{chiang2024intercellular}.

The dynamics of topological defects is also informed by geometrical frustration with fascinating implications \cite{Vafa2024}. To this end, the application of the multi-phase-field model on curved surfaces \cite{Happel2024} can help decipher the role of local curvature in disrupting the local order and its biological consequences (Fig. \ref{fig2}f), often manifested through the local stress fields as discussed next.  

\subsection{Active stress chains and force transmission}
The spatiotemporal dynamics of mechanical stresses play an important but poorly understood role in gene expression \cite{Weinstein2019}, genomic damage and potential mutation \cite{Gupta2022}, dynamics of expanding populations \cite{Schreck2023} as well as cell differentiation \cite{Zhang2022}. At the same time, the transition of mechanical information is a primary suspect for facilitating collective self-organization, on a length scale much larger than an individual cell, crucial for development and regeneration \cite{JasonGao2016,Vining2017}. A recent computational study investigated solid-like to liquid-like transition in active monolayers by focusing on active stress chains and emerging stress patterns \cite{monfared2024short} (Fig. \ref{fig2}g). By mapping this transition onto the 2D random percolation universality class using two independent active derives, this work revealed the short-range nature of stress correlation near this transition, providing a new context for understanding the non-equilibrium physics of active systems and its connections to glass, jamming and two-dimensional melting transitions. Recent research has shown that activity-induced annealing can lead to a ductile-to-brittle transition in amorphous solids, which parallels the behavior observed in biological tissues under mechanical stress \cite{Sharma2025}. This insight is crucial for understanding how mechanical forces influence tissue integrity and failure. In Section \ref{section:4}, focused on biological physics, we will thoroughly discuss the role of force transmission on biological processes such as the fate of an extruding cell and cell competition. 


\FloatBarrier
\section{Biological physics of cells: integration of experiments with multi-phase-field modeling}\label{section:4}
On top of understanding the physical properties of active matter, multi-phase-field models have demonstrated the ability to capture the emergence of various biological phenomena driven by the mechanical properties of cells. For instance, cancer cells are known to differ mechanically from healthy counterparts. By modeling a single cancer cell in a layer of `normal' cells, the 2D multi-phase-field model demonstrated that elasticity mismatch alone is sufficient to significantly increase the motility of the cancer cell \cite{Palmieri2015}. Additionally, simulations of extensile monolayers can capture a spontaneous formation of gaps (Fig. \ref{fig2}h) -- a phenomenon observed in epithelial monolayers \cite{zheng2017epithelial}. Finally, it is possible to model mixtures of both active and passive phases, hence, reproducing interactions between cells and their environment, such as extracellular matrix (ECM) (Fig. \ref{fig2}a) \cite{MoreiraSoares2020} or blood vessels.

In this section, we will focus on the role of multi-phase-field models on the interdisciplinary advances in multicellular systems, starting from parametrization of 2D cell monolayers to 3D models of embryogenesis. We focus particularly on problems where multi-phase-field modeling has been combined and studied together with relevant biological experiments. To bridge these different scales, we first discuss the importance of quantitative parametrization before examining the emergent behaviors that arise from mechanical interactions.

\subsection{Quantitative modeling: model parameters informed by experiments}
The precise matching of physical model parameters to biological systems is an outstanding challenge. Existing cell-based models have qualitatively reproduced flow fields and mechanical stresses around topological defects~\cite{balasubramaniam2021investigating}, as well as captured the amplitude and period of collective oscillations in confined epithelial monolayers~\cite{Peyret2019}. Using the cell length, cell velocity, and force units from experiments, the simulation length, time, and force units can be mapped directly into the physical units of $\delta L \sim 2\mu m$, $\delta t \sim 0.1 min$, and $\delta F \sim 200 nN$, respectively (Table \ref{tab:params}). 

Beyond direct parameter extraction, a powerful approach is to transform model parameters into dimensionless groups using Buckingham-Pi theorem~\cite{sonin2004generalization,marquez2023biological} (Table \ref{tab:disk}). The results of sensitivity analyses of dimensionless parameters in both 2D \cite{mueller2019emergence, Peyret2019, balasubramaniam2021investigating} and 3D \cite{Balasubramaniam2025} showed that the ratio of active to elastic forces, $(\zeta R_0)/\gamma$, is the predominant factor in governing cell deformation. In the 3D formulation, the ratio of the cell-cell to cell-substrate adhesion, $\omega_{\text{cc}}/\omega_{\text{cs}}$, is an additional significant parameter that affects collective cell motion in monolayers \cite{monfared2023mechanical}. These results set the foundation for understanding how mechanical interactions translate into emergent collective behaviors.
\begin{table*}[t] \label{tab:params}
\resizebox{\textwidth}{!}{
    \begin{tabular}{ | c | c | c | c | c | }
        \hline 
        Simulation parameter & Physical meaning & Numerical value &  Mapping to physical units & Measured physical value\\
        \hline \hline
        $R_0$ & initial cell radius & $8$ & $15\mu m$ & $10-20\mu m$~[this study] \\
        \hline
        $\gamma$ & cortex tension & $0.008-0.016$ & $800-1600pN/\mu m$ & $1000-2000pN/\mu m$~\cite{chugh2017actin,taneja2020precise} \\
        \hline
        $\xi$ & friction coefficient & $1$ & $600nN.s/\mu m$ & $O(10^{2})nN.s/\mu m$~\cite{arciero2011continuum,cochet2014border} \\
        \hline
        $\omega_{\text{cc}}$ & cell-cell adhesion force & $0.006-0.012$ & $12-24nN$ & $O(10^{1}-10^{2})nN$~\cite{priest2023characterizing} \\
        \hline
        $\omega_{\text{cs}}$ & cell-substrate adhesion force & $0.001-0.002$ & $2-4nN$ & $O(10^{1}-10^{2})nN$~\cite{trichet2012evidence,tan2020regulation} \\
        \hline
        $\zeta$ & active stress & $0.00001-0.001$ & $0.5-50Pa$ & $O(10^{0}-10^{1})Pa$~[this study] \\
        \hline
        $\tau_{\text{pol.}}$ & cell polarity alignment time & $200$ & $20min$ & $O(10^{1} min)$~\cite{peyret2019sustained} \\
        \hline
        $\alpha$ & single cell traction magnitude & $0.05$ & $10nN$ & $1-30nN$~\cite{du2005force} \\
        \hline
        $\kappa_{\text{cc}}$ & cell-cell repulsion force & $0.5$ & $--$ & $--$ \\
        \hline
        $\kappa_{\text{cs}}$ & cell-substrate repulsion force & $0.15$ & $--$ & $--$ \\
        \hline
        $\mu$ & stiffness of volume constraint & $45$ & $--$ & $--$ \\
        \hline
        $\lambda$ & width of diffuse interface & $1.5$ & $--$ & $--$ \\
        \hline
    \end{tabular}
    }
    \caption{Mapping of model parameters to physical units. `$--$' indicates that mapping to physical units is non-applicable, since the parameter is model-specific, and no experimental measurement is available. Note that in~\cite{arciero2011continuum,cochet2014border} the friction coefficient is mapped by using $Pa$ units for force, giving friction dimensions of $nN.s/\mu m^{3}$, and is here converted to $nN.s/\mu m$ using the cell size as the relevant length scale. Adapted from~\cite{Balasubramaniam2025}.}
    \label{tab:disk}
\end{table*}

\begin{table*}[t]
    \resizebox{\textwidth}{!}{
    \begin{tabular}{ | c | c | c | }
        \hline 
        Dimensionless parameter & Physical interpretation & Sensitivity analyses \\
        \hline \hline
        $\lambda/R_{0}$  & diffuse interface width compared to cell radius & not-sensitive:\\
        & & the width of the diffuse interface\\
        & & is set smaller than the cell radius ($\lambda/R_{0} \ll 1$)\\
        \hline
        $\kappa R_0^2/\mu$  & cell-cell overlap to compressibility ratio & not-sensitive:\\
        & & only needed for keeping cell integrity\\
        & & and avoiding overlaps between phase-fields \\
        \hline
        $\kappa_{\text{cc}}/\kappa_{\text{cs}}$  & cell-cell to cell-substrate repulsion energy ratio & not-sensitive:\\
        & & only needed to avoid overlap between phase-fields\\
        & & representing cells and substrate \\
        \hline
        $\omega_{\text{cc}}/\omega_{\text{cs}}$  & cell-cell to cell-substrate adhesion energy ratio & one of the main control parameters \\
        \hline
        $\zeta R_0/\gamma$  & contractility to stiffness ratio & one of the main control parameters \\
        \hline
        $\alpha\tau_{\text{pol.}}/(\xi R_0)$  & ratio of realignment to directed motion time & not-sensitive: drives flocking behavior \\
        \hline
    \end{tabular}
    }
    \caption{Dimensionless model parameters and their physical interpretation. Adapted from~\cite{Balasubramaniam2025}.}
    \label{tab:disk}
\end{table*}
\subsection{From micro- to macro-scale}
Mechanical interactions govern the collective behavior of tissues, impact protein distributions, and have been shown to induce various biological functions \cite{Ladoux2017}. When placed on a solid substrate, an epithelial cell exerts contractile forces -- a pair of equal and opposite forces acting inwards along the cellular axis. However, at the collective cell level, the monolayers yield extensile behavior, as observed from the flow fields and stress patterns around topological defects \cite{sawtopological2017, blanch2018turbulent, kawaguchi2017topological}. 

\begin{figure*}[tbp]
 \centering
\includegraphics[width=1\linewidth]{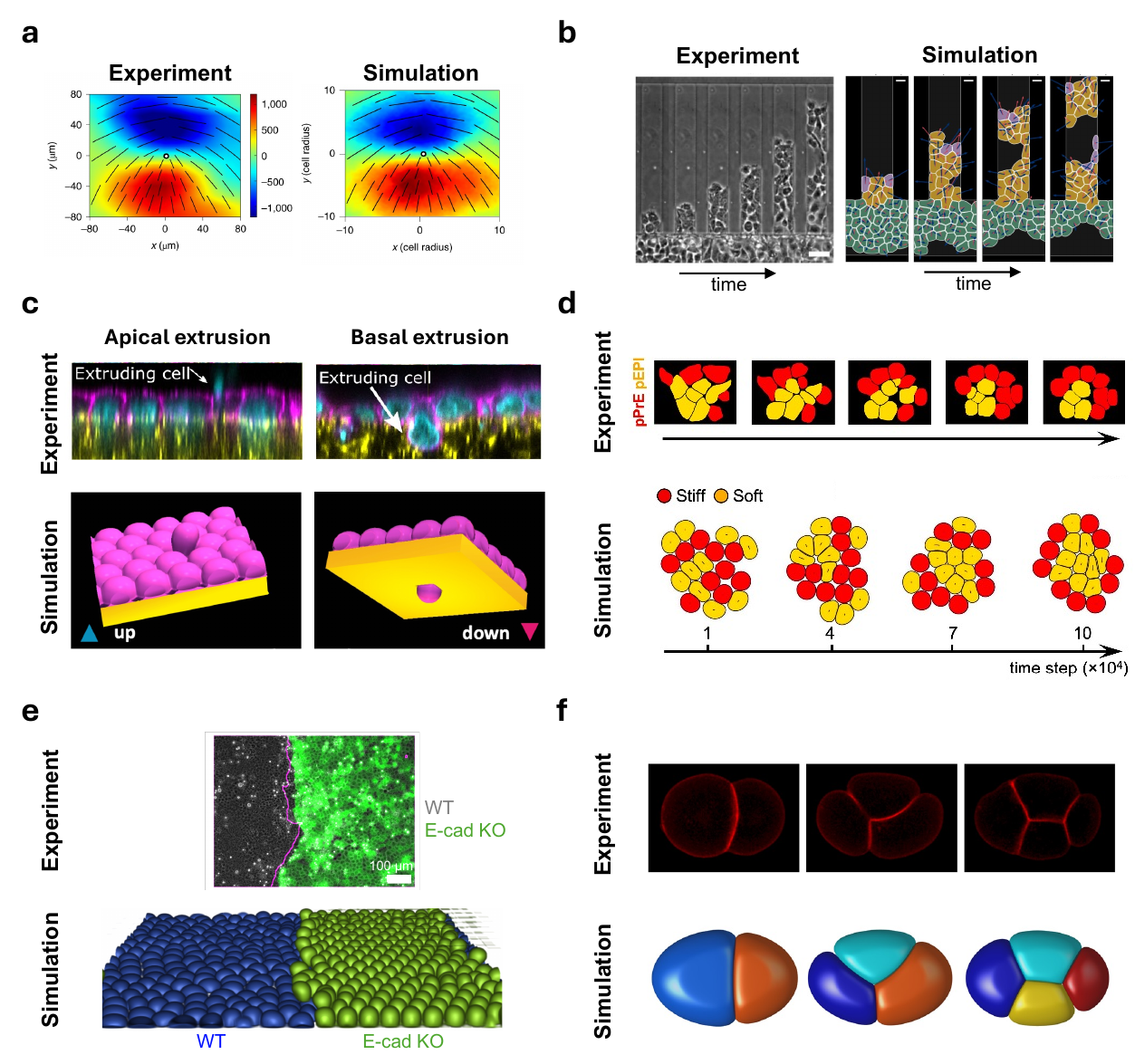}
\caption{\label{fig4} {\bf Biological physics of cells: integration of experiments with multi-phase-field modeling.} a) Average isotropic stress around a +1/2 defect obtained from experiments on E-cadherin knock-out MDCK cells (left) and simulations (right). Adapted from \cite{balasubramaniam2021investigating}. b) Left: phasecontrast snapshots of invading human A431 epidermoid carcinoma cells confined in 50 $\mu$m microchannels. Right: evolution of rupture for simulations of cells in microchannels with 50 $\mu$m width. Adapted from \cite{wang2024confinement}. c) Top: orthogonal view of immunostaining of MDCK WT (left) and MDCK E-cad KO (right) monolayers grown on 2D type I collagen gels, actin (magenta), collagen (yellow) and nuclei (cyan). Arrows indicate extruding cell. Bottom: snapshots from simulations demonstrating basal and apical extrusions by varying activity and cell-cell adhesion. Adapted from \cite{Balasubramaniam2025}. d) Top: time-series snapshots of experimental segregation (pEPI cells in yellow, pPrE cells in red). Bottom: simulation snapshots of the segregation of a 20-cell aggregate of soft (yellow) and stiff (red) cells. Adapted from \cite{ritter2025}. e) Mechanical cell competition between two colliding assays of WT and E-cad KO cells, displaying the cell type and location of extrusions; white spots in experimental image frame (top) and simulations (bottom). Adapted from \cite{schoenit2024force}. f) Comparison between \textit{ab initio} simulations and \textit{in vivo} fluorescence images of \textit{C. elegans} embryonic morphologies from 2- to 4-cell stages. Adapted from \cite{Kuang2023}.}
\end{figure*}

This puzzling difference between micro- and macro-scale properties of cell monolayers has been explained using a 2D multi-phase-field model, which simultaneously accounts for the properties of single cells and yields the resulting collective dynamics. The model accurately captured isotropic stresses around topological defects (Fig. \ref{fig4}a) and predicted that the reduction in the strength of cell-cell interactions leads to an increase in cell-substrate interactions, which results in extensile nematic dynamics \cite{balasubramaniam2021investigating}. This proposed mechanism was confirmed in experiments showing increase in the strength of cell-substrate adhesion and confirming the emergence of intercellular extensile stresses through laser ablation experiments~\cite{balasubramaniam2021investigating}.

A complementary explanation was proposed by Zhang et al.~\cite{zhang2023active}, who demonstrated that fluctuations in phase-field models, rather than mean-field interactions alone, can also drive the observed extensile behavior. This highlights the need for a more comprehensive approach that integrates both deterministic and stochastic contributions to multi phase-field dynamics.

\subsection{Experimentally-informed modeling of collective cell migration in 2D}
Collective cell migration \textit{in vivo} is a complex and dynamic process essential for wound healing, embryogenesis, and cancer invasion \cite{lauffenburger1996cell}. One aspect of this complexity is the influence of confinement and environmental interactions, which shape the collective and single-cell migration strategies observed in biological systems. Multi-phase-field models have provided powerful tools to study these constraints, successfully replicating the dissociation of clusters from collectively invading cancer cells in confined geometries \cite{wang2024confinement}. By explicitly linking self-propulsion strength to chemical gradients and geometric confinement, such models have demonstrated that a solid-to-liquid transition, along with the presence of leader cells exhibiting enhanced motility, are both necessary for rupture events to occur (Fig. \ref{fig4}b). Additionally, they have revealed how parameters such as channel width and cell-cell adhesion influence the likelihood of cluster rupture, shedding light on migration strategies across diverse microenvironments.  

Beyond confinement effects, recent studies have emphasized another layer of complexity in cell migration—interactions with self-deposited traces. Experimental work has shown that migrating cells can modify their own microenvironment by secreting extracellular matrix components or depleting substrate adhesiveness, thereby altering the effective landscape for subsequent migration \cite{d2021cell}. Multi-phase-field models incorporating such feedback mechanisms reveal emergent behaviors where cells exhibit persistent or oscillatory motion in response to their own tracks~\cite{perez2024deposited}, highlighting a form of self-guided navigation that can either enhance or inhibit migration efficiency. These findings suggest that cell motility is not only dictated by external constraints but also by the dynamic remodeling of the environment through cellular activity, underscoring the importance of coupling biochemical and mechanical interactions in predictive models of migration.  

By integrating insights from both confinement-induced migration and self-trace interactions, multi-phase-field models provide a mechanistic framework to explore emergent migration behaviors. Such approaches offer a deeper understanding of how cells navigate complex landscapes, with implications for tissue development, immune responses, and metastatic invasion.

\subsection{From 2D to 3D: mechanical imprints of live versus dead cell elimination}
Even though in appropriate limits cell monolayers can be approximated as a two-dimensional problem~\cite{chiang2024multiphase}, there are various biological phenomena that require consideration of the third dimension. This is particularly relevant in cell extrusion -- the process by which cells are removed from a monolayer. 

Cells can be extruded either live or dead. Determining how the fate of an extruding cell is determined remains a major question in biology with significant implications for tissue and organ development under both normal and pathological conditions. In a combined experimental and multi-phase-field modeling approach, it was shown that weakening cell-cell contacts, effectively altering the force transmission capability of cells, affects the local stress patterns near an extruding cell \cite{Balasubramaniam2025}. The multi-phase-field model was able to capture the distributions for the local stress fields and their temporal evolutions for both normal and transformed cell types. These quantitative insights revealed how the intensity and duration of local stress fields shape the survival or death of an extruding cell. 

Furthermore, the same multi-phase-field model was employed to predict the direction of cell extrusion, apical vs. basal, on a soft porous collagen substrate. The extended model incorporated irregular environments through an additional phase-field representing a deformable extracellular matrix. The model predicted that cells with lower cell-cell adhesion and higher contractility preferentially extrude basally, consistent with experiments (Fig. \ref{fig4}c). These insights have since been extended to 3D cysts in matrigel, reinforcing the universality of mechanical cues governing the fate of an extruding cell. 

\subsection{Heterogeneous cell populations} Biological tissues are not perfectly uniform in terms of cell mechanical properties and often exhibit heterogeneity, where cells with different mechanical properties coexist. One of the advantages of multi-phase-field models is their ability to incorporate such heterogeneity and isolate the effects of specific mechanical parameters.

The cell-sorting phenomenon has been of particular interest \cite{graham2024cell}(Fig.~\ref{fig2}i). A 2D multi-phase-field model demonstrated that differences in elasticity alone can drive autonomous segregation, where softer cells, which collectively exhibit fluid-like behavior, become surrounded by stiffer cells, which collectively exhibit solid-like behavior \cite{ritter2025}. This mechanism explains the segregation of primitive endoderm (pPrE) and epiblast (pEPI) cells observed in mouse embryonic stem cell experiments (Fig. \ref{fig4}d).

Additional phase-field studies on cell sorting, such as \cite{graham2024cell} and \cite{balasubramaniam2021investigating}, have highlighted the role of adhesion heterogeneity and active forces in determining sorting outcomes. In particular, recent research has demonstrated that mixtures of cells exhibiting distinct dipolar activities -- either extensile or contractile -- can spontaneously segregate into elongated domains. This segregation is driven by differences in cellular diffusivity, akin to the behavior of Brownian particles connected to thermostats at varying temperatures. Notably, this mechanism operates independently of traditional thermodynamic factors, underscoring the significance of active forces in cellular self-organization. Additionally, phase-field modeling combined with experiments on MDCK cells demonstrated how the interplay between cell-substrate interactions and active stresses regulate sorting of extensile and contractile cells when they adhere strongly to the substrate \cite{balasubramaniam2021investigating}.

The emergence of new physics through multi-phase-field modeling has been particularly enlightening. These models have revealed that the interplay between active forces and mechanical properties can lead to novel segregation patterns, such as the formation of extensile and contractile domains. This insight is crucial for understanding how cells self-organize in a tissue, providing a more comprehensive framework for studying developmental processes and disease progression.

Overall, these modeling predictions and comparison with experiments~\cite{balasubramaniam2021investigating} highlight the distinct mechanisms of phase separation driven by differences in cellular activity compared to traditional differential adhesion or line tension models. While initial studies focused on cadherin-mediated surface tension as a primary driver of tissue segregation~\cite{maitre2012adhesion}, it is now evident that intercellular adhesion is not the sole factor. Theoretical models suggest that a combination of cell surface tension and contractility can also drive cell sorting~\cite{manning2010coaction}. Although differential adhesion and line tension may still play a role, the results emphasize the critical importance of cell-substrate interactions and intracellular stresses in regulating cell sorting within strongly adherent cellular monolayers~\cite{niessen2002cadherin,balasubramaniam2021investigating}.

Hence, incorporating cell heterogeneity at various levels could provide a more comprehensive framework for understanding self-organization in developing tissues.

\subsection{Mechanical competition in heterogeneous cell populations}
While competition within heterogeneous cell populations can dictate survival dynamics at the tissue level, the spatial organization of these populations is also influenced by external constraints and geometric confinement. Cells in heterogeneous populations not only coexist and segregate but also compete with each other for space and nutrients. Cell competition is crucial not only for maintaining homeostasis and fighting pathogens but also for the progression of diseases, such as cancer \cite{van2023cell}. By modeling two distinct cell types (Fig. \ref{fig4}e) and measuring directly the forces the cells exhibit on each other, the multi-phase-field model demonstrated the emergence of in-plane stress fluctuations at the interface of two competing cell populations \cite{schoenit2024force}. This was found to be a crucial feature in understanding the outcome of mechanical cell competition: the cell type with a stronger cell-cell adhesion strength can more effectively transmit this interfacial stress fluctuations away from the frontline cells near the interface onto the bulk. This translates into a competitive advantage relative to cells with a weaker cell-cell adhesion strength, and hence stress transmission capability, where in-plane stress fluctuations become localized and induce out-of-plane stresses - reminiscent of the Poisson effect in elasticity - which can lead to a higher probability for cell elimination. Importantly, these predictions were compared with and confirmed by direct experimental measurements of forces in between competing cell types~\cite{schoenit2024force}.

The multi-phase-field modeling approach is pivotal in revealing the intricate physics of mechanical competition. It allows for the detailed simulation of differential force transmission capabilities between cell types, which play a crucial role in determining the outcome of cell competition. Cells with stronger intercellular adhesion exhibit higher resistance to elimination due to their ability to efficiently transmit forces to neighboring cells, thereby maintaining tissue integrity and boundaries. This efficient force transmission results in increased mechanical activity at the interface of competing cell populations, characterized by large stress fluctuations. These fluctuations can induce upward forces, leading to the elimination of cells with weaker adhesion properties.

Moreover, the multi-phase-field model underscores the importance of mechanical forces in maintaining tissue homeostasis and preventing pathological conditions such as tumorigenesis. The ability of cells to mechanically outcompete each other through directed migration, crowding, and differences in growth rates underscores the complex interplay between biochemical and mechanical factors in cell competition. Understanding these mechanisms through multi-phase-field modeling provides valuable insights into the development of therapeutic strategies aimed at modulating cell competition to treat diseases characterized by abnormal cell proliferation and invasion.

\subsection{Confined cellular systems in 3D}
Extending beyond competition within planar tissues, the role of confinement becomes particularly relevant in three-dimensional contexts, where cells are not only influenced by their neighbors but also by surrounding structural constraints. In many biological conditions, cells do not simply exist as monolayers. Instead, they form three-dimensional structures that are confined and may change over time due to proliferation or external environmental factors. This is particularly important during morphogenesis, when an organism starts as a single cell, and undergoes division and self-organization -- all dictated by genetic factors, as well as physical forces \cite{heisenberg2013forces}. The phase-field formalism allows for the prescription of various geometries and, with proper parameterization using experimental data, can be suitably adapted for unraveling the physics behind the early stages of morphogenesis. By extracting accurate geometrical constraints and relevant parameter values from experimental three-dimensional time-lapse cellular \textit{in vivo} imaging of developing \textit{C. elegans}, the model can precisely reproduce the time evolution of the location and shapes of every cell during the morphogenetic transformation from 1-- to 4--cell stages (Fig. \ref{fig4}f) \cite{kuang2022computable}. The model predicts how physical factors, such as cell division timing, cell division orientation, and cell-cell attraction matrix, govern robust morphological evolution at 6--, 7--, and 8--cell stages.

The examples discussed illustrate the versatility of multi-phase-field models in capturing key biophysical processes across different scales in biological tissues. These models have proven essential in elucidating cell migration, collective behavior, extrusion dynamics, and mechanical competition. However, significant challenges remain, which we discuss in the next section.

\section{Challenges and future directions}
Despite the significant progress made, several challenges remain in the development and application of multi-phase-field models. One major challenge is the accurate representation of the complex mechanical and biochemical interactions within tissues. While current models capture key mechanical properties at a single-cell level, many features are yet to be incorporated.

Mechanotransduction, the process by which cells convert mechanical stimuli into biochemical signals, is a critical area for future research. Phase-field models can be extended to include mechanotransduction pathways, allowing for a more comprehensive understanding of how mechanical forces influence cell behavior and tissue development \cite{romani2021crosstalk}. This will involve integrating mechanical feedback mechanisms and signaling pathways into the models to simulate how cells sense and respond to their mechanical environment.

Modeling the mechanical interaction between the cell nucleus and the active cytoskeleton is another important direction. The nucleus plays a crucial role in cellular mechanics, and its interaction with the cytoskeleton affects various cellular processes, including migration and differentiation \cite{Camley2014,Moure2020,chojowski2024role}. Incorporating the mechanical properties of the nucleus and its interactions with the cytoskeleton into phase-field models will provide deeper insights into the role of nucleus mechanics in cell behavior.

Using phase-field models to understand and explore emergent collective modes and long-range ordering of cells is another exciting direction. Cells in tissues exhibit collective behaviors that arise from their interactions and mechanical properties. Phase-field models can be used to study these emergent behaviors and understand the principles that govern long-range ordering in cell collectives \cite{najem2016phase}. In this vein, another promising direction is the coupling of multi-phase-field models with hydrodynamic flows. Recent research has revealed the importance of water transport, viscosity, and flow in cellular processes. Water and hydraulic pressure play essential roles in cell shape changes, motility, and tissue function, generating significant mechanical forces \cite{li2020importance}. Coupling multi-phase-field models with hydrodynamic models will allow for a more accurate representation of these processes, providing insights into how fluid dynamics influence cell behavior and tissue mechanics. This approach can help elucidate the role of hydraulic resistance and external hydraulic pressures in cell polarization and motility, as well as the impact of fluid-structure interactions on cellular and tissue dynamics \cite{shelley2024flows}. In addition, extending multi-phase-field models to capture interstitial fluid transport and viscoelastic properties of the extracellular matrix will enhance their applicability to realistic tissue environments, further bridging the gap between simulations and experimental observations.

\begin{figure*}[t!]
 \centering
\includegraphics[width=1\linewidth]{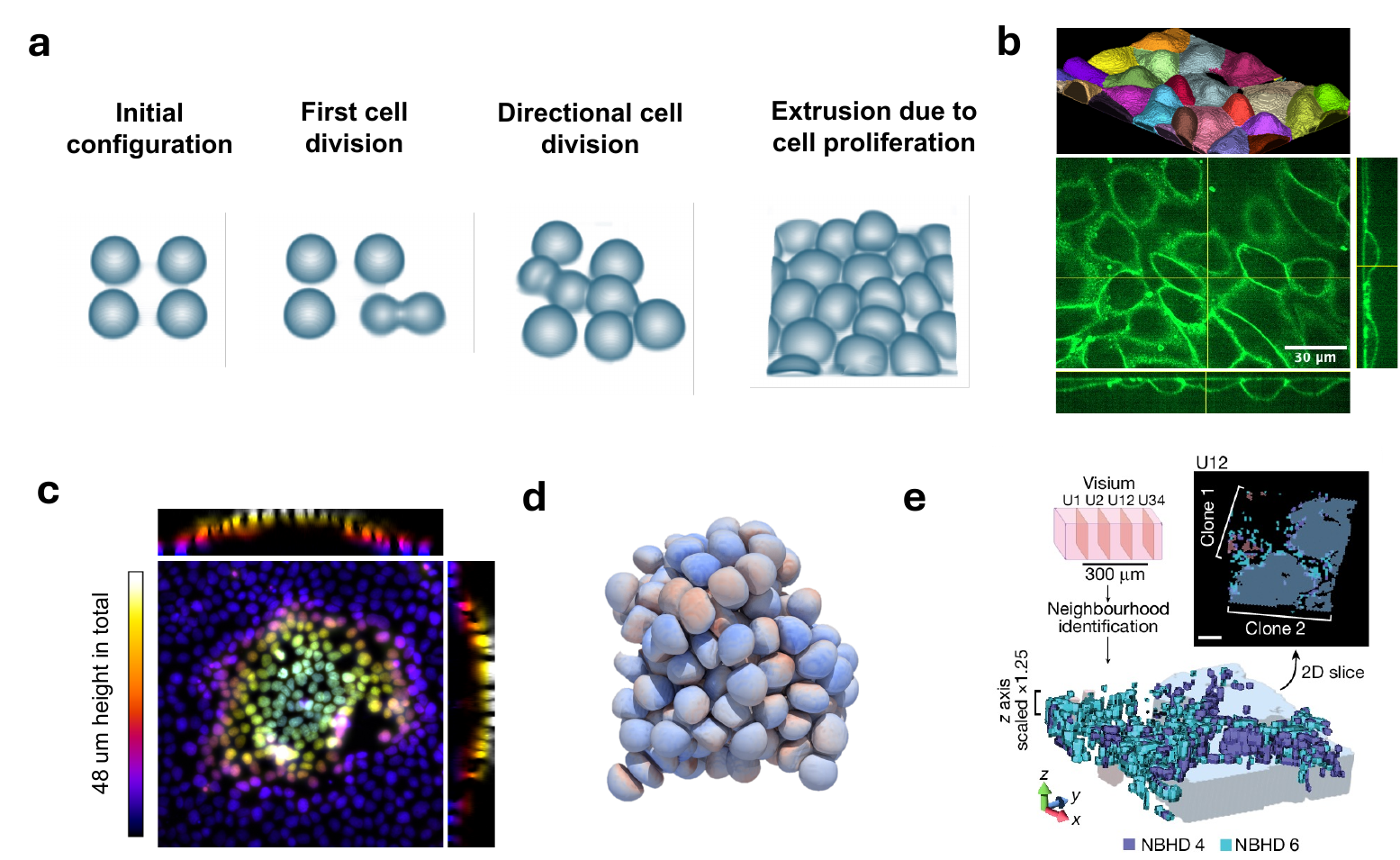}
\caption{\label{fig5} {\bf Challenges and future directions for multi-phase-field modeling of biological tissues.} a) Snapshots of a simulation demonstrating proliferating cells in a three-dimensional multi-phase-field model. b) Bottom: orthoviews (xy, xz, yz) of MDCK cells that are stained with a membrane dye. Top: 3D Segmentation of the cell surface performed using Cellpose. Images courtesy of Valeriia Grudtsyna. c) MDCK Ecadherin-RFP cells that have been cultured on a glass coated with fibronectin for about 2 days and then fixed and dyed with Hoechst. Images courtesy of Valeriia Grudtsyna. d) Snapshot of multi-phase-field simulation of a 3D cell cluster. e) A reconstructed 3D neighbourhood volumes of a metastasis tissue sample. Adapted from \cite{mo2024tumour}. }
\end{figure*}
\FloatBarrier
An important future challenge is the explicit incorporation of different types of fluctuations — thermal, active, and biochemical — into phase-field models. While thermal fluctuations play a minor role in large-scale tissue mechanics, they remain relevant at the subcellular level. Active fluctuations, arising from cytoskeletal remodeling and stochastic motor activity, introduce non-equilibrium noise that significantly affects cellular dynamics, phase separation, and collective behavior. Incorporating such fluctuations into multi-phase-field models is crucial for understanding how cell-to-cell variability, stochastic biochemical signaling, and active force generation contribute to emergent tissue properties. Furthermore, biochemical noise in reaction-diffusion systems can lead to spatial heterogeneities and stochastic transitions in cell fate decisions, requiring models to integrate stochasticity in both mechanical and biochemical fields. Addressing these challenges will be key to improving the predictive power of phase-field models and their ability to capture the full complexity of living tissues.

Capturing shape changes of cells in collectives is another critical area for future research. Cells in tissues undergo complex shape changes during migration, division, and differentiation. Extending phase-field models to accurately simulate these shape changes, including cell proliferation as demonstrated in Fig. \ref{fig5}a, will provide valuable insights into the mechanics of tissue morphogenesis and the factors that drive these shape changes \cite{link2024modelling}. Advanced modeling techniques, such as the cellular Potts model, have shown promise in accurately predicting 3D cell shapes in structured environments \cite{link2024modelling}.

Volume fluctuations and the compressibility of cells in collectives are also important aspects to consider. Cells in tissues exhibit volume changes due to various factors, including osmotic pressure and mechanical forces, as clearly shown in the image of MDCK cell monolayers Fig. \ref{fig5}b. Modeling these volume fluctuations and their impact on collective cell behavior will enhance our understanding of tissue dynamics and the mechanical properties of cell collectives \cite{zehnder2015cell}. Recent studies have shown that cell volume fluctuations can significantly influence collective migration and tissue organization, highlighting the need for more detailed models that capture these dynamics \cite{jipp2024active}.

Extending phase-field models to account for complex 3D structures is crucial, as even 2D monolayers may exhibit heterogeneities in the z-direction, as shown by the dome formation in MDCK cell monolayers (Fig. \ref{fig5}c). Furthermore, the modeling of organoids and spheroids within a multi-phase-field framework is a promising direction for understanding tissue and organ development, as these systems mimic the structure and function of tissues and organs. Phase-field models can be used to simulate the growth and development of these systems (Fig. \ref{fig5}d), providing insights into the factors that influence their morphology and function \cite{tanida2024predicting}. Recent studies have demonstrated the potential of phase-field models to predict organoid morphology and understand the mechanical factors that drive their self-organization \cite{tanida2024predicting}. However, a comprehensive framework should also account for chemical signaling and reaction-diffusion processes that regulate organoid development. This can be achieved by coupling mechanics with biochemical activity, such as through a Flory–Huggins type free energy functional, to model phase separation and cellular differentiation \cite{Zwicker2016,Bauermann2022}. Expanding phase-field models to integrate such chemical interactions will enhance our understanding of how biochemical gradients and mechanical forces coordinate organoid growth, patterning, and fate decisions.

Moreover, recent advances in the 3D imaging of patient tumors and identifying cancer cell subpopulations (Fig. \ref{fig5}e) pave the way toward understanding the interplay between the mechanical properties of cells and the tumor microenvironment. Despite their success in capturing collective organization of cells, multi-phase-field models face several challenges in fully representing the complexity of tumor dynamics. One key challenge is incorporating the heterogeneous mechanical properties of tumors, where stiffness gradients and local deformations influence cell migration and invasion. Additionally, improving the resolution and computational efficiency of these models remains an ongoing challenge, particularly when simulating large-scale tumors with intricate mechanical interactions. Another crucial direction is the integration of poroelastic effects to account for the role of interstitial flows and pressure gradients, which significantly impact tumor expansion and cell motility. Finally, advancing these models to include the dynamic remodeling of the extracellular matrix and its feedback on tumor progression will be essential for capturing the evolving mechanical landscape of growing tumors. Addressing these challenges will enhance the predictive power of multi-phase-field models and enable more precise insights into tumor mechanics and morphology.

Overall, multi-phase-field models continue to be a cornerstone of biological and physical research, offering unparalleled insights into the dynamic and complex nature of cellular, tissue, and physical systems. The integration of diverse biological processes and mechanical interactions into these models has enabled researchers to explore the fundamental principles governing tissue dynamics and development, paving the way for new discoveries and advancements in the field. Future expansions of these models, incorporating biochemical signaling, hydrodynamic interactions, and tissue-specific regulatory mechanisms, will be essential for developing a more unified and predictive framework for biological morphogenesis. Additionally, systematically incorporating stochastic effects and active fluctuations into these models will provide a deeper understanding of robustness in developmental processes and the role of noise in cellular decision-making. The application of physical principles in these models not only enhances our understanding of biological systems but also drives innovation in computational methods, leading to more accurate and predictive simulations of cellular and tissue dynamics.

\section*{Acknowledgments}
We apologize to our colleagues whose valuable work could not be cited in this review due to focusing on recent advancements on only multi-phase-field models of cell layers.

\noindent A.A. acknowledges support from the EU’s Horizon Europe research and innovation program under the Marie Sklodowska-Curie grant agreement No. 101063870 (TopCellComm). A.D. acknowledges funding from the Novo Nordisk Foundation (grant no. NNF18SA0035142 and NERD grant no. NNF21OC0068687), Villum Fonden Grant no. 29476, and the European Union via the ERC-Starting Grant PhysCoMeT, grant no. 101041418.

\bibliographystyle{unsrt}
\bibliography{main}
\end{document}